\documentclass[12pt]{article}
\input epsf.sty
\newcommand{\tl}{\wt\lambda}

\newcommand{\HH}{{\cal H}}

\newcommand{\LL}{{\cal L}}

\newcommand{\wt}{\widetilde}
\newcommand{\wh}{\widehat}

\newcommand{\bd}{\bar{\rm D}}

\newcommand{\TT}{{\cal T}}
\newcommand{\dt}{(\vec \nabla T)^2}
\newcommand{\gsim}{{>\atop \sim}}
\newcommand{\lsim}{{<\atop \sim}}

\newcommand{\be}{\begin{equation}}
\newcommand{\ee}{\end{equation}}
\newcommand{\ben}{\begin{eqnarray}\displaystyle}
\newcommand{\een}{\end{eqnarray}}
\newcommand{\refb}[1]{(\ref{#1})}
\newcommand{\p}{\partial}
\newcommand{\sectiono}[1]{\section{#1}\setcounter{equation}{0}}

\begin{document}
{}~
\hfill\vbox{\hbox{hep-th/0209122}
}\break

\vskip .6cm

\centerline{\Large \bf
Time and Tachyon}

\medskip

\vspace*{4.0ex}

\centerline{\large \rm
Ashoke Sen }

\vspace*{4.0ex}

\centerline{\large \it Harish-Chandra Research
Institute}

\centerline{\large \it  Chhatnag Road, Jhusi,
Allahabad 211019, INDIA}

\centerline {and}
\centerline{\large \it Abdus Salam International Centre for Theoretical
Physics}

\centerline{\large \it Strada Costiera 11, 34014, Trieste, Italy}

\centerline{E-mail: asen@thwgs.cern.ch, sen@mri.ernet.in}

\vspace*{5.0ex}

\centerline{\bf Abstract} \bigskip

Recent analysis suggests that the classical dynamics of a tachyon on an
unstable D-brane is described by a scalar Born-Infeld type action with a
runaway potential. The classical configurations in this theory at late
time are in one to one correspondence with the configuration of a system
of non-interacting (incoherent), non-rotating dust.  We discuss some
aspects of canonical quantization of this field theory coupled to
gravity, and explore, following earlier work on this subject, the
possibility of using the scalar field (tachyon) as the definition of
time in quantum cosmology. At late `time' we can identify a subsector in
which the scalar field decouples from gravity and we recover the usual
Wheeler - de Witt equation of quantum gravity.

\vfill \eject

\baselineskip=16pt

\tableofcontents

\sectiono{Introduction and Summary} \label{s1}

Recent studies in time dependent solutions involving the
tachyon\cite{0203211,0203265,0207105} (see also
\cite{0202210,0205085,0205098,0206102,0207107,0204203,0207004,0208028})
indicates
that open string
field
theory on
a
non-BPS D-brane or a
brane-antibrane pair has time dependent, spatially homogeneous classical
solutions of arbitrary low energy density
measured from the tachyon vacuum.
Whereas the energy
density associated with these solutions remains constant in
time due to the
usual conservation law,
the pressure goes
to zero asymptotically.
This, in turn, suggests a specific form of the low energy
effective action describing
the dynamics of the tachyon near
the minimum of the tachyon potential.
If in particular we consider the
dynamics of the tachyon on a space-filling D-brane system, then this low
energy effective action will be an action in (9+1) or (25+1) dimensions in
superstring and bosonic string theories respectively, and has the form
\be \label{ei1}
S = -\int d^{p+1} x V(T) \sqrt{1 + \eta^{\mu\nu} \p_\mu T \p_\nu T}\, ,
\ee
where $p=9$ for type IIA or IIB superstring theories and $25$
for the bosonic string
theory.
(Independent of this analysis, this form of the
action has been suggested from other
considerations\cite{effective,0003221}.)
We are using the convention $\alpha'=\hbar=c=1$ and
$\eta^{\mu\nu}=\hbox{diag}(-1,1,\ldots 1)$.
The tachyon potential $V(T)$ has a maximum at $T=0$, and behaves for large
$T$ as $e^{-\alpha T/2}$
where $\alpha=1$ for bosonic string theory and $\sqrt 2$ for the
superstring theory\cite{0204143}.\footnote{This apparently differs from
the potential computed in boundary string field theory\cite{bsfttach}, but
as
has been argued in \cite{0208142}, this could simply be an effect of
field
redefinition which includes derivative terms.} Thus the minimum of the
potential is at $T=\infty$.
Unlike in the case of ordinary scalar field, there
is no physical scalar particle
associated with the perturbative quantization
of the tachyon field $T$ around the minimum of its potential\cite{0204143},
consistent with the conjecture put forward in \cite{soliton}.

The space-time independent classical
solutions in this effective field theory
include solutions with fixed energy density and asymptotically vanishing
pressure, exactly as we see in explicit string calculation.
On the other hand this
effective field theory also contains other inhomogeneous
classical solutions, and at late time these solutions are
in one to one correspondence with configurations of
non-interacting, non-rotating
dust\cite{0204143,0208019}.\footnote{Various
other aspects of the action
\refb{ei1}
have been
discussed recently in \cite{0208217,0209034}.}
These inhomogeneous solutions of the effective field
theory can also be identified
with specific classical
solutions in string field theory on
space-filling D-brane system as follows.
First of all, beginning with sufficient number of space filling
D-branes or brane
anti-brane pairs, we can construct, as classical
solutions, an arbitrary
distribution of D0-$\bd0$-brane pairs (for IIA string theory) or non-BPS
D0-branes (for IIB or bosonic string theory) \cite{soliton,solitonb}.
We can
now
repeat the construction of time
dependent solutions
with arbitrary energy
on each of these 0-branes. This gives a distribution of point objects, each
carrying
arbitrary
energy, and by choosing an appropriate distribution of these objects with
appropriate energies, we can construct a configuration of arbitrary energy
density and zero pressure, exactly as we obtain from the classical
solutions of the
field theory described by the action \refb{ei1}.
Thus it would seem that the effective action \refb{ei1} captures
at least to some
extent the classical dynamics of open string
field theory around the tachyon
vacuum.

In this paper we consider coupling of the action \refb{ei1} to
gravity, and discuss
some aspects of the quantum theory describing this coupled
system.\footnote{Much work has been done recently on
(semi-)classical analysis of
this coupled system, keeping in mind possible application to
cosmology\cite{cosmo}.}
Before we
summarize the results, let us make a few cautionary remarks:
\begin{enumerate}
\item The tachyon effective action \refb{ei1} has not been
derived
from first principles; it is only consistent with the time evolution of
the energy-momentum tensor of various
classical solutions in string theory at
late time.

\item Although the energy momentum tensor associated with the
above-mentioned classical
solutions in string theory approach finite limit asymptotically,
the
sources for some massive closed string states created by time-like
oscillators acting on the vacuum grow exponentially with
time\cite{0208196,0208142}.\footnote{A different
kind of exponential growth, -- that
of the mass of the open string states
on the original D-brane, -- has been discussed
recently in \cite{0209090}. The relationship of this to the results of
\cite{0208196,0208142} is also not clear.} It is
possible to add appropriate coupling of these massive closed
string fields to the tachyon $T$ appearing in \refb{ei1} to reproduce
these results; however the physical
meaning of this exponential growth is not completely clear at present.
In our analysis we shall only include the coupling of
\refb{ei1} to gravity.
Some comments on the possible role
of these exponentially growing terms
in the boundary state will be made in
section \ref{s4}.

\item Even if the effective action \refb{ei1} describes correctly the
classical
tachyon dynamics,
quantizing this classical field
theory coupled to gravity may not describe correctly the physics in
quantum string
theory.
\end{enumerate}
Nevertheless, we take the point of view that the
classical field theory described by the action \refb{ei1}
is an
interesting field theory, and studying quantum properties in this theory
is an interesting problem in its own right. It is in this spirit that the
results of this paper should be interpreted.\footnote{Of course \refb{ei1}
coupled to gravity is a non-renormalizable field theory, and so much of
the analysis will have to be at a formal level. If this
analysis turns out to be relevant for string theory then the ultraviolet
divergence will be automatically regularized by string theory. } We shall
continue to refer to the scalar field $T$ as tachyon for convenience.

The main results of our analysis are summarized below:
\begin{enumerate}
\item As already mentioned, at the classical level solutions of the
equations of motion of the field theory described in \refb{ei1} at late
time are in one to one
correspondence with configuration of non-rotating, non-interacting dust.
Furthermore at late time (large $x^0$) the `classical vacuum' solution of
the equations of motion approach $T\simeq x^0$. This makes $T$ a candidate
for describing time at this classical level.
We construct a more general class of field
theories, of which the action
\refb{ei1} is a special case, all of which have the same property. This
class of field theories have been analyzed independently in
\cite{0208217,0209034} in the context of tachyon dynamics.
(See also
\cite{9904075} for related studies.)

\item As is well-known, there is no natural notion of intrinsic time in
classical general relativity, and this leads to a number of conceptual
difficulties in formulating a quantm theory of gravity. (See
\cite{rov,kutc,9409001,0108040,9210011}, references therein, and
citations
to
these
papers.) It has been argued earlier\cite{kutc,9409001} that
some of these
problems can be resolved in a theory with dust coupled to gravity, where
one uses dust world-lines to foliate space-time, and takes the proper
time along the world-line passing through any given space-time point as
the definition of time at that point. Since the classical low energy
configurations of the
tachyon field theory can be represented as non-interacting,
non-rotating dust,
we could use
this to give a definition of time. We show that this procedure leads to
the identification of the value of the tachyon field at a given space-time
point as the time coordinate associated to this point.  Given this natural
definition of time, one can rewrite the Hamiltonian constraints on the
wave-functional (the Wheeler - de Witt equations) so that they take the
form of many fingured time Schrodinger equation.

\item
At late `time', {\it i.e.} for large values of the field $T$,
our results are
identical to that of \cite{kutc,9409001} for non-rotating, incoherent
dust coupled to gravity. In particular
at late `time' the quantum theory of the tachyon coupled to
gravity has a subsector where the solution to the Wheeler - de Witt
equation is
independent of
$T$. Hence
the
tachyon field decouples from gravity. and we recover the usual
vacuum gravity.
However this does not happen at early `time', {\it i.e.}
for small or finite values of
$T$, and the solution to the Wheeler - de Witt equation necessarily has a
non-trivial dependence on $T$ in this region.
This shows that in this theory
we are inevitably led to
`time' dependent wave-functionals of the universe.

\end{enumerate}

The rest of the paper is organised as follows. In section \ref{s2} we
review the classical low energy effective action describing the tachyon
dynamics, discuss the region of validity of this effective action and
(approximate) conservation laws following from this effective action. We
also argue, based on classical dynamics of the tachyon effective action,
that it is natural to identify the tachyon as time coordinate. In section
\ref{s3} we couple the tachyon effective action to gravity and analyze the
constraints of general relativity for this coupled system. In particular,
following \cite{9409001} we show how the Hamiltonian constraints of
general
relativity can be simplified for this system, and the Wheeler - de Witt
equations can be rewritten as many fingured time Schrodinger equation.
We conclude the paper in section \ref{s4} with a few comments on various
issues related to
this paper.

\sectiono{Tachyon Without Gravity} \label{s2}

We begin
this section by reviewing the salient features of the classical
field theory describing the tachyon field without coupling to
gravity.

\subsection{The action}
The field theory proposed in
\cite{0203265,0204143}
describing the dynamics of the tachyon near the minimum of its potential
is given by:
\ben \label{ez1}
S &=& \int d^{p+1} x \, \LL\, , \nonumber \\
\LL &=& -  V(T) \, \sqrt{1 + \eta^{\mu\nu}\p_\mu T
\p_\nu T} \, = \, - V(T) \, \sqrt{-\det A}
\, ,
\een
where
\be \label{ey2}
A_{\mu\nu} = \eta_{\mu\nu} + \p_\mu T \p_\nu T\, ,
\ee
and
\be \label{ezz1}
V(T) \simeq e^{-\alpha T/2}  \quad \hbox{for large $T$}\, ,
\ee
with
\ben \label{e2}
\alpha &=& 1 \qquad \hbox{for bosonic string theory} \nonumber \\
&=& \sqrt 2 \qquad \hbox{for superstring theory}\, .
\een
The form of the potential $V(T)$ was derived by
requiring that if we let the tachyon
roll down from its maximum, then at late time the pressure associated with
the configuration falls off as $K_0 \exp(-\alpha x^0)$ for some constant
$K_0$. This result, in turn, is derived from a stringy analysis of the
boundary state associated with the rolling tachyon solution.
Unless stated otherwise,
in the rest of the analysis we shall take $(p+1)$ to be equal to the
dimension
of space-time (10 for superstring and 26 for bosonic string theory), so
that \refb{ez1} describes
the dynamics of the tachyon on a space-filling brane system.

The energy momentum tensor computed
from the action \refb{ez1} is given by:
\be \label{etens}
T_{\mu\nu} = {V(T) \, \p_\mu T \p_\nu T \over \sqrt{1 + \eta^{\rho\sigma}
\p_\sigma T \p_\rho T}} - V(T)\,
\eta_{\mu\nu}\, \sqrt{1 + \eta^{\rho\sigma}
\p_\rho
T \p_\sigma T}\, .
\ee

\subsection{Hamiltonian Formulation and Classical solutions}

The solutions to the equations of motion described by the action
\refb{ez1} can be found easily
by
working in
the hamiltonian
formalism\cite{0009061,9901159,9704051,0010240,0204143,0209034}.
Defining the momentum conjugate to $T$ as:
\be \label{ex1}
\Pi(x) = {\delta S \over \delta (\p_0 T(x))} = {V(T) \p_0 T\over \sqrt{1
- (\p_0 T)^2 + \dt}}\, ,
\ee
we can construct the Hamiltonian $H$:
\be \label{ex2}
H = \int d^p x \, (\Pi \p_0 T - \LL) \equiv \int d^p x \HH, \qquad
\HH = T_{00} = \sqrt{\Pi^2 + (V(T))^2 } \, \sqrt{1 + \dt}
\, .
\ee
The equations of motion derived from this hamiltonian take the form:
\be \label{ex4p}
\p_0 \Pi(x)= -{\delta H\over \delta T(x)} = \p_j \bigg( \sqrt{\Pi^2 +
V^2}
\, { \p_j
T
\over
\sqrt{1 + \dt} }\bigg) - {V(T) V'(T) \over \sqrt{\Pi^2 + V^2}} \,
\sqrt{1 + \dt}\, ,
\ee
\be \label{ex5p}
\p_0 T(x) = {\delta H \over \delta \Pi(x)} = {\Pi\over
\sqrt{\Pi^2 + V^2}}\, \sqrt{1 + \dt} \, .
\ee
In the limit of large $T$ ({\it i.e.}
near the tachyon vacuum) at fixed $\Pi$,
we can
ignore the
$V^2\simeq e^{-\alpha T}$ term, and the Hamiltonian and the equations
of motion take the form:
\be \label{eham}
H = \int d^p x \, |\Pi| \sqrt{1 + \dt}\, ,
\ee
\be \label{ex4}
\p_0 \Pi(x)= \p_j \bigg( |\Pi|\, { \p_j
T
\over
\sqrt{1 + \dt} }\bigg)\, ,
\ee
\be \label{ex5}
\p_0 T(x) = {\Pi\over
|\Pi|}\, \sqrt{1 + \dt} \, .
\ee
{}From \refb{ex5}, we see that
in this limit we have $(\p_0T)^2 - \dt= 1$.

These equations can be rewritten in a suggestive form by
defining\cite{0204143}
\be \label{eadd1}
u_\mu \equiv -\p_\mu T, \qquad \epsilon(x) \equiv |\Pi(x)|/
\sqrt{1 + \dt}\, .
\ee
Eqs.\refb{ex4}, \refb{ex5} then take the form:
\be \label{eadd2}
\eta^{\mu\nu} u_\mu u_\nu = -1, \qquad \p_\mu (\epsilon(x) u^\mu) = 0\, .
\ee
Expressed in terms of these new variables, $T_{\mu\nu}$ given in
\refb{etens} take the form:
\be \label{eadd3}
T_{\mu\nu} = \epsilon(x) u_\mu u_\nu\, ,
\ee
where we have used the small $V(T)$ approximation
and used the equations of
motion \refb{ex4}, \refb{ex5}.
These are precisely the equations governing the motion of non-rotating,
non-interacting dust,
with $u_\mu$ interpreted as the local $(p+1)$-velocity
vector\cite{0204143}, and
$\epsilon(x)$ interpreted as the local rest mass density.
Conversely, any configuration describing flow of non-rotating,
non-interacting dust can
be interpreted as a solution of the equations of motion \refb{ex4},
\refb{ex5}.
\subsection{Tachyon as time}

Particular solutions of the equations
\refb{ex4}, \refb{ex5} can be obtained
by taking
\be \label{ex6}
\Pi(x) = f(\vec x)\, , \qquad T(x) = x^0\,,
\ee
where $f(\vec x)$ is any arbitrary positive definite function of the
spatial coordinates.
The energy density associated with such a solution is proportinal to
$f(\vec x)$. If we further take $f(\vec x)$ to be independent of $\vec x$,
then we get a spatially uniform energy density.

Solution \refb{ex6} naturally leads to a definition of the {\it classical
vacuum} solution. Since the hamiltonian is positive semi-definite,
the classical vacuum must have zero energy
density. However we shall define this by a limiting procedure.
If the
tachyon begins rolling at some large value of $T$ with a small
potential
energy $\varepsilon$, then no matter how small $\varepsilon$ is,
as the tachyon rolls down the potential hill towards large $T$, the
system will eventually settle down to the configuration:
\be \label{efconf}
\Pi \simeq \varepsilon, \qquad  T \simeq x^0 - {1\over \alpha} \, \ln({2
\alpha C\varepsilon^2}) + C e^{-\alpha x^0},
\qquad V(T) \simeq \varepsilon \, \sqrt{2 \alpha C} \, e^{-\alpha x^0/2}
\ee
for some constant $C$,
so that the
energy
density $V(T)/\sqrt{1 - (\p_0 T)^2}$ remains constant at $\varepsilon$,
and approaches
$\Pi/\p_0 T$ as $T\to\infty$\cite{0203265}.
We now define the classical
vacuum solution as the $x^0\to\infty$,
$\varepsilon\to
0^+$ limit of this configuration. In this limit:
\be \label{evac1}
\Pi\to 0^+, \qquad \p_0 T \to 1, \qquad V(T) / \Pi \to 0\, .
\ee

For the vacuum solution $T(x)=x^0$ at late time. This indicates that it is
natural to take $T(x)$ as the definition of the time coordinate in
general.
This definition of cosmic time agrees with a
proposal of \cite{kutc,9409001}
for defining the time coordinate in general
relativity by coupling it to a system of incoherent, non-rotating
dust. According to
this proposal, we foliate the space-time by world-lines of dust, and the
proper time along a dust world-line
is to be interpreted as the time coordinate
of various space-time points lying along that world-line. In our case,
according to \refb{eadd1}, the
local $(p+1)$-velocity of the dust is given by $u_\mu = -\p_\mu T$. Thus
the proper time $\tau$ measured along a dust world-line satisfies:
\be \label{ezz6}
d\tau = - u_\mu dx^\mu = d T\, .
\ee
Hence we see that we can identify $\tau$ with $T$. Thus $T$ measures the
proper time along any of the dust world-lines
and hence can be identified
with the cosmic time coordinate associated with any space-time point.

This analysis also shows that for any
$V(T)/\Pi>0$, however small, if we go
sufficiently far back in time then the
time dependence of the solution becomes
non-trivial and $T=x^0$ is no longer a
valid solution. Thus the identification
of $T(x)$ as time breaks down at early time. We shall encounter a quantum
version of this phenomenon in the next section.

\subsection{Lorentz transformation properties}

It is easy to derive the Lorentz transformation property of
$\Pi$.
Since $\p_\mu T$ transforms as a
covariant 4-vector, and $\eta^{\mu\nu} \p_\mu T \p_\nu T$ transforms as a
scalar, we see, from eq.\refb{ex5p} that $\sqrt{
(\Pi^2+V^2) / (1 + \dt)}$
transforms as a scalar, and $({\Pi \sqrt{1+\dt
}/ \sqrt{\Pi^2+V^2}}, \p_i T)$ transforms as a covariant
4-vector. Thus
\be \label{e4vec}
\left(\Pi, {\sqrt{\Pi^2 + V^2}\, \p_i T\over \sqrt{1 + \dt}}\right)\, ,
\ee
transforms as a covariant 4-vector. In the limit of large $T$ where $V$
vanishes, we see that
\be \label{e4vec1}
\left(\Pi, {|\Pi| \, \p_i T\over \sqrt{1 + \dt}}\right)
\ee
transforms as a covariant 4-vector, and $|\Pi| / \sqrt{1 + \dt}$
transforms as a scalar. This is
consistent with the interpretation of $|\Pi| /
\sqrt{1 + \dt}$ as
the local rest mass density, and of \refb{e4vec1} as
the local $(p+1)$-momentum density of the dust.

\subsection{Approximate conservation laws}

Let us work in the region $|\Pi|>> V(T)$, so that the equations
\refb{eham}-\refb{ex5} give valid approximation to the classical
dynamics.\footnote{Since $T\simeq x^0$, this requires the
energy density contained in the tachyon field to be large compared to
$e^{-\alpha x^0/2}$. If we take $x^0$ to be the present age of the
universe, then $e^{-\alpha x^0/2}$ is an incredibly small number, and an
energy density of this order will be undetectable for all practical
purpose.} In this approximation the dynamics of the system has a peculiar
property.
Eq.\refb{ex5} shows that
\be \label{e2.1}
|\p_0 T| \ge 1\, .
\ee
Continuity of $T$ in space-time will then imply that
if some region in space-time has positive (negative) $\p_0 T$,
then all space-time points must
have positive (negative) $\p_0 T$.
Eq.\refb{ex5} also shows that
the sign of $\p_0
T$ is the same as the sign of $\Pi$. Hence
$\Pi(x)$ must
either be
positive everywhere in space-time or negative everywhere in
space-time.\footnote{This argument of course is somewhat circular, since we
have assumed that $|\Pi|> V(T)$ to begin with. What we want to emphasize here
is
that once we have arrived at \refb{eham}-\refb{ex5} by taking this limit,
then we can forget about the restriction $|\Pi|> V(T)$, and still
configurations which are continuous in space-time must have $\Pi(x)$ positive
or
negative
everywhere. See section \ref{s4} for more discussion on this.
We also note that for a generic initial condition, time evolution of $T$
typically leads to cautics\cite{0208019} where the
approximation leading to the original action \refb{ez1} itself breaks
down. Hence all our classical analysis will be applicable only in those
regions of
space-time which are free from these singularities.
}
Since the effect of the potential $V(T)\propto \exp(-\alpha
T/2)$ is to drive $\Pi$ towards positive value, for studying late time
dynamics of the system we can restrict $\Pi(x)$ to be positive
everywhere in space-time as long as the field
configuration remains non-singular. Eqs.\refb{eham}-\refb{ex5} can then be
rewritten as
\be \label{ehams}
H = \int d^p x \, \Pi \sqrt{1 + \dt}\, ,
\ee
\be \label{ex4s}
\p_0 \Pi(x)= \p_j \bigg( \Pi(x)\, { \p_j
T
\over
\sqrt{1 + \dt} }\bigg)\, ,
\ee
\be \label{ex5s}
\p_0 T(x) = \sqrt{1 + \dt} \, .
\ee
The local rest mass density $\epsilon(x)$ is now given by
$\Pi(x)/ \sqrt{1
+ \dt}$, and preservation of the condition $\Pi>0$ by the equations of
motion \refb{ex4s}, \refb{ex5s} is simply the statement that the
local rest mass density of the dust remains positive during the time
evolution.
Consistency of this restriction on $\Pi$ is also confirmed by noting from
\refb{e4vec1} that $\Pi(x)$ is the time component of a time-like
vector field, and hence the condition $\Pi(x)>0$ is Lorentz invariant.

In this limit, the quantity,
\be \label{ezz3}
\Pi_0(x^0) \equiv \int d^p x \, \Pi(x) \, ,
\ee
is conserved. Indeed,
We see from \refb{ehams} that the Poisson bracket of $H$ and $\Pi_0$
vanishes:
\be \label{ezz4}
\{ H, \Pi_0\} = 0\, .
\ee

\subsection{Region of validity of \refb{ez1}}

Of course, \refb{ez1} cannot be the complete answer for the tachyon
effective
action in string theory; it can only be a good approximation in some
limit.
This effective action ignores all quantum corrections, and
even at the classical level the form of this action has not been derived
from first principles.
We conjecture that this is a good approximation to
the full classical effective
action if the following criteria are satisfied:
\begin{enumerate}
\item The tachyon field $T$ must be large. Only in this limit we expect
the lagrangian density to have a factorized form as given in \refb{ez1}.

\item The second and higher derivatives of the tachyon must be
small. In particular even when $T$ is
large, we do not expect
\refb{ez1} (or more generally \refb{eac1})
to be a valid description of the
dynamics of the tachyon unless
the second and higher derivatives of the tachyon are small. This can be
seen as follows. For uniform tachyon field, \refb{etens} gives:
\be \label{etmn}
T_{00} = V(T)/\sqrt{1 - (\p_0 T)^2}\, , \qquad T_{ij} =
-V(T) \, \sqrt{1 - (\p_0 T)^2}\, \delta_{ij} \, , \qquad T_{i0}=0\, .
\ee
Thus the
energy-momentum tensor of the system at an instant when $\p_0 T=0$ must be
of the
form $-V(T) \, \eta_{\mu\nu}$. This in particular implies that the
pressure
$p$
is equal to the negative of the energy density $\rho$ at this instant.
This result also holds for the more general action \refb{eac1}
discussed later.
This
however does not agree with
the exact stringy formula for the evolution of $p$
and
$\rho$ calculated from string theory\cite{0203265}:\footnote{Here we are
quoting the results for bosonic string theory, but the results for
superstring theory lead to a similar conclusion.}
\be \label{ezz2}
\rho(x^0) = {1\over 2}\, \TT_p\, (1 + \cos(2\tl\pi)), \qquad
p(x^0) = - \TT_p\, \bigg[ {1\over 1 + \sin(\tl\pi) e^{x^0}} +
{1\over 1 + \sin(\tl\pi) e^{-x^0}} - 1 \bigg]\, ,
\ee
where $\TT_p=V(0)$ is the energy density (tension)
of the static unstable configuration
at the top
of the tachyon potential, and $\tl$ is a parameter labelling the total
energy density of the system.
{}From this we see that at $x^0=0$, where the
tachyon begins rolling with $\p_0 T=0$, the ratio $p/\rho$ does not
approach $-1$ even for
$\tl\simeq 1/2$ which corresponds to low energy density, and hence
large
initial value of the tachyon
field $T$. This in turn implies that the higher derivative terms must be
contributing to the effective action. Only at
late time, when $\p_0 T$ approaches 1, and hence the second and higher
derivatives of
$T$ go to zero, the effective action \refb{ez1} is valid, and
gives $p\propto
e^{-\alpha x^0}$.
For $\alpha=1$ this corrctly reproduces \refb{ezz2} for large $x^0$.

Note that the constraint that the second and
higher derivatives of the tachyon must be small implies that
this description is expected to be valid as long as $\p_\mu u_\nu$ is
small in magnitude (in string units), where $u_\mu=-\p_\mu T$ is the local
velocity vector of the dust. As pointed out in
\cite{0208019}, however,
actual
classical evolution of the equations of motion beginning with a generic
inhomogeneous tachyon field configuration runs into the
problem that $\p_\mu
u_\nu$ becomes large during the course of evolution, and hence
we will need to invoke
stringy modification of the dynamics to follow the evolution of the
system near these
points.

\item We also expect that the description of the system by a low energy
effective action will be meaningful only when the energy density of the
system is small. This will require $V(T)/\sqrt{1-(\p_0 T)^2}$ to be small.

\end{enumerate}

\subsection{Other Lagrangians leading to the same limiting Hamiltonian}

Before we conclude this section, we would like to point out that the
limiting form \refb{eham} of the Hamiltonian can in fact be obtained from
a
more general class of Lagrangians than considered here.
(This analysis has been done
independently in \cite{0209034}, see also \cite{0208217}.) In fact,
based on the analysis in boundary string field theory, a different form of
the action for describing the classical tachyon dynamics was proposed in
refs.\cite{0205085,0205098}. The action
\refb{ez1}, as well as the action discussed in \cite{0205085,0205098},
can be thought of as special cases of a general class of actions of the
form:
\be \label{eac1}
S = - \int d^{p+1} x \, V(T) \, F(\eta^{\mu\nu}\p_\mu T
\p_\nu T) \, ,
\ee
where $V$ and $F$ are two functions with the property that $V(T)$ goes to
zero as $T\to\infty$ from positive side, and $F(u)$ and its derivatives
are well defined in the range $-1 < u \le 0$, and has a singularity at
$u=-1$ where $F'(u)\to +\infty$, and $F(u)/F'(u)$ vanishes. (The
location of
the singularity of $F(u)$
can be changed by a rescaling of the field $T$; we have chosen a
convenient normalization so that the singularity is at $-1$.)
The
energy-momentum tensor computed from this action is given by:
\be \label{eac2}
T_{\rho\sigma}= 2 V(T) F'(\eta^{\mu\nu}\p_\mu T \p_\nu T) \p_\rho T
\p_\sigma T - V(T) F(\eta^{\mu\nu}\p_\mu T \p_\nu T) \eta_{\rho\sigma}\, .
\ee
For spatially uniform tachyon field $T_{00}$ must be conserved. Since as
$T$ rolls towards infinity,
$V(T)\to 0$, in order to keep $T_{00}$ fixed $\p_0 T$ must approach its
critical value 1 where $F'(\eta^{\mu\nu}\p_\mu T \p_\nu T)$ blows up.
Eq.\refb{eac2} then shows that the pressure, being proportional to $-V(T)
F(\eta^{\mu\nu}\p_\mu T \p_\nu T)\propto - T_{00} F(\eta^{\mu\nu}\p_\mu T
\p_\nu T) / F'(\eta^{\mu\nu}\p_\mu T \p_\nu T)$ in this limit, vanishes.
By adjusting the form of $V(T)$ for large $T$ and the behaviour of the
function $F(u)$ near $u\simeq -1$ we can ensure that the pressure falls
off as $\exp(-\alpha x^0)$ for large $x^0$.

The
momentum
conjugate to $T$, computed from the action \refb{eac1}, is given by:
\be \label{eac3}
\Pi = 2 V(T)\, \p_0 T \, F'(\eta^{\mu\nu}\p_\mu T \p_\nu T)\, .
\ee
The Hamiltonian is given by:
\be \label{eac4a}
H\equiv \int d^p x \, \HH(x) \, ,
\ee
where
\be \label{eac4b}
\HH(x) = T_{00}(x) = 2 V(T) F'(\eta^{\mu\nu} \p_\mu T \p_\nu T) (\p_0 T)^2
+ V(T) F(\eta^{\mu\nu} \p_\mu T \p_\nu T)\, .
\ee
For $\eta^{\mu\nu}\p_\mu T \p_\nu T\simeq
-1$, we have
\be \label{eac4c}
\p_0 T = \pm \sqrt{1 + \dt}\, ,
\ee
and the first term in \refb{eac4b} dominates the second one.
Using \refb{eac3} and \refb{eac4c} we now get:
\be \label{eac4}
\HH \simeq |\Pi| \sqrt{1 + \dt}\, ,
\ee
which is the same as the one given in \refb{eham}.
Thus the equations of motion near this critical point are identical to
those given in \refb{ex4}, \refb{ex5}. The energy momentum tensor,
expressed in terms of $\Pi(x)$ also has the same form as
given in eqs.\refb{eadd1}-\refb{eadd3} in this limit,
since the second term on the right hand side
of \refb{eac2} can be neglected.
Thus the system near its critical point describe non-rotating,
non-interacting dust exactly
as the system analyzed earlier.
For explicit analysis we shall continue to work with the action
\refb{ez1}-\refb{ezz1}, but many of our results will be valid also for
this more
general class of field theories.

\sectiono{Coupling Tachyon Effective Action to Gravity} \label{s3}

The
coupling of
the tachyon effective action \refb{ez1} to supergravity fields (including
the
world-volume gauge and fermionic fields on the unstable D-brane) has been
described in
\cite{9909062,0003221,0208142,0206102}.
In this section we restrict ourselves to an analysis of the
coupling of \refb{ez1}
to
gravitational field,
discuss how it modifies the constraints of general relativity, and rewrite
the Wheeler - de Witt equation for this
coupled system as many fingured time
Schodinger equation. We follow closely the analysis of \cite{9409001}.

\subsection{The action}

For simplicity we shall
work with the action \refb{ez1}, but identical results can be derived for
the general action \refb{eac1} near the critical point where $T$ is large
and $g^{\mu\nu}\p_\mu T \p_\nu T$ is close to $-1$. If $g_{\mu\nu}$
denotes the space-time metric,
then the action \refb{ez1} coupled to gravity takes the form:
\be \label{egr1}
S = - \int d^{p+1} x \, V(T) \, \sqrt{-\det g} \, \sqrt{1 +
g^{\mu\nu}\p_\mu T
\p_\nu T}\, .
\ee
We now consider foliation of space-time by space-like hypersurfaces
$\Sigma$ parametrized by coordinates $\xi^a$ ($1\le a\le p$) and label the
different leaves of foliation by the parameter $t$. Then the space-time
coordinates
$x^\mu$ can be thought of as functions of $(t, \xi^a)$. $\p_a x^\mu$ for
$a=1,\ldots p$) are the $p$ vectors tangential to $\Sigma$. We define by
$n^\mu$ the unit time-like vector normal to $\Sigma$:
\be \label{egr2}
n^\mu g_{\mu\nu} \p_a x^\nu = 0, \qquad g_{\mu\nu} n^\mu n^\nu = -1\, .
\ee
We also define the lapse function $N^\perp$ and the shift vector $N^a$
through the decomposition:
\be \label{egr3}
\p_t x^\mu = N^\perp n^\mu + N^a \p_a x^\mu\, ,
\ee
and denote by
\be \label{egr5}
h_{ab} = g_{\mu\nu} \p_a x^\mu \p_b x^\nu
\ee
the induced metric on $\Sigma$, and by $h^{ab}$ the matrix
inverse of $h_{ab}$.
All information about the metric $g_{\mu\nu}$ is contained in $h_{ab}$,
$N^a$ and $N^\perp$.
We have
\be \label{egr4}
g^{\mu\nu} = - n^\mu n^\nu + h^{ab} \p_a x^\mu \p_b x^\nu\, , \qquad
\sqrt{-\det g} \, d^{p+1}x = N^\perp \, \sqrt{\det h} \, dt \, d^p\xi\, .
\ee
Using these relations we can rewrite \refb{egr1} as
\ben \label{egr6}
S &=&  \int dt \int_\Sigma d^p\xi \, \LL\, , \nonumber \\
\LL &=& - N^\perp \sqrt{\det h} \, V(T) \sqrt{1 - (N^\perp)^{-2} (\p_t T -
N^a
\p_a T) (\p_t T - N^b \p_b T) + h^{ab}\p_a T \p_b T }\, , \nonumber \\
\een
where $\p_t T = \p_t x^\mu \p_\mu T$ and $\p_a T = \p_a x^\mu \p_\mu T$.

\subsection{The Hamiltonian and the constraints}

{}From \refb{egr6} we can compute the momentum $\Pi(t,\xi)$
conjugate to $T$ as
\be \label{egr7}
\Pi(t,\vec\xi) = {\delta S\over \delta(\p_t T)}
= {\sqrt{\det h} \, V(T) (N^\perp)^{-1} (\p_t T - N^c \p_c T) \over
\sqrt{1 - (N^\perp)^{-2} (\p_t T - N^a
\p_a T) (\p_t T - N^b \p_b T) + h^{ab}\p_a T \p_b T }}\, .
\ee
The Hamiltonian associated with the tachyon
effective action in the background
gravitational field is given
by:
\be \label{egr8}
H^T = \int d^p \xi \, ( \Pi(t, \vec\xi) \p_t T - \LL)
= \int d^p \xi \, (N^\perp \HH^T_\perp + N^a \HH^T_a)\, ,
\ee
where
\be \label{egr9}
\HH^T_\perp = \sqrt{\Pi^2 + V(T)^2 \det h } \, \sqrt{ 1 + h^{ab} \p_a T
\p_b
T}\, , \qquad
\HH^T_a = \Pi \, \p_a T\, .
\ee
The full hamiltonian, when we treat the metric also as a dynamical
variable, and include the standard Einstein-Hilbert term in the action,
is
given by
\be \label{egrfull}
H^{total}
= \int d^p \xi \, [N^\perp (\HH^T_\perp + \HH^G_\perp)
+ N^a (\HH^T_a+\HH^G_a)] \, + \, \hbox{boundary terms}\, ,
\ee
where $\HH^G_\perp$ and $\HH^G_a$ are the
contributions to $\HH_\perp$ and $\HH_a$
from the Einstein's action:
\be \label{egr12}
\HH^G_\perp = G_{abcd} p^{ab} p^{cd} - \sqrt{\det h} \, R(h)\, , \qquad
\HH^G_a = - 2 D_b p^b_a\, .
\ee
Here $p^{ab}$ is the momentum conjugate
to $h_{ab}$, $R(h)$ is the Ricci scalar on
$\Sigma$ with metric $h_{ab}$, $D_a$ denotes
covariant derivative along $\Sigma$, and
\be \label{egr13}
G_{abcd} = {1\over 2} (\det h)^{-1/2} \, (h_{ac} h_{bd} + h_{ad} h_{bc} -
{2\over p-1} \, h_{ab} h_{cd})\, .
\ee
The boundary terms appearing in \refb{egrfull}
have been discussed in
\cite{reggteit}. In our
analysis we shall not be careful in keeping track of these
terms. For a discussion of boundary terms for non-interacting,
non-rotating
dust coupled to gravity see, for example, ref.\cite{9811062}. We can
include in $H^G_\perp$ and $H^G_a$ the contribution from any other
standard matter field coupled to gravity
without affecting the rest of the
analysis.

The constraints of general relativity can now be written as
\be \label{egr10}
\HH^T_\perp + \HH^G_\perp = 0\, ,
\ee
and
\be \label{egr11}
\HH^T_a + \HH^G_a = 0\, .
\ee
Since from \refb{egr9} we see that $H^T_\perp$ is positive, \refb{egr10}
forces $H^G_\perp$ to be negative.

For large $T$, we can ignore the $V(T)$
term in eq.\refb{egr9}. The hamiltonian
equations of motion for $T$ then gives
\be \label{eteqn}
\p_0 T = {\Pi\over |\Pi|}\, N^\perp\,
\sqrt{1 + h^{ab} \p_a T \p_b T}\, + N^a \p_a
T\, .
\ee
This, together with \refb{egr3}, \refb{egr4},
gives $g^{\mu\nu} \p_\mu T \p_\nu T =
-1$. As in section \ref{s2},
this allows us to restrict $\p_\mu T$
to be a future pointing time-like vector
field.\footnote{We are using the convention that either for covariant or a
contravariant vector, a future pointing
time-like vector will have its time
component positive in a locally inertial frame of reference.
In this convention if
$n^\mu$ is a future pointing time-like vector,
then $-n_\mu$ will be a future
pointing time-like vector. \label{f1}} On the
other hand, using \refb{egr3},
\refb{egr7}, we
see that
$\Pi$ is given by a positive quantity
multiplying $n^\mu \p_\mu T$. This, being an
inner product of a future pointing contravriant
vector $n^\mu$ and a future pointing
covariant vector $\p_\mu T$ (see footnote
\ref{f1} for convention) is positive
definite. This allows us to
restrict $\Pi(\vec\xi, t)$ to be positive.
Using \refb{egr9}
we can now write the hamiltonian constraint \refb{egr10} in this
approximation as
\be \label{eexcon}
\Pi + (1 + h^{ab} \p_a T \p_b T)^{-1/2} \HH^G_\perp = 0\, .
\ee
We can also rewrite this using
\refb{egr11}
as\cite{9409001}
\be \label{egr14}
\HH_\uparrow \equiv \Pi + H^G_\uparrow =0, \qquad
H^G_\uparrow\equiv -\sqrt{ (\HH^G_\perp)^2 -
h^{ab} \HH^G_a
\HH^G_b} =0 \, ,
\ee
where we take the positive square root of
$\big((\HH^G_\perp)^2 -
h^{ab} \HH^G_a
\HH^G_b\big)$ in \refb{egr14} so that
$H^G_\uparrow$ is negative. In this form, $\Pi$
appears linearly in this constraint, and there is no $T$-dependence.

\subsection{Constraint algebra}

The algebra of the costraints $\HH_\uparrow$ and $\HH_a$
can be easily computed.
In particular one can show, following \cite{9409001}, that
\ben \label{enewcom}
\{ \HH^G_\uparrow(t,\vec\xi), \HH^G_\uparrow(t,\vec\xi')\} &=& 0\, ,
\nonumber \\
\{ \HH^G_\uparrow(t,\vec\xi), \HH^G_a(t, \vec\xi')\} &=&
\HH^G_\uparrow(t, \vec\xi')
\p_a \delta(\vec\xi
-\vec\xi')\, , \nonumber \\
\{ \HH^G_a(t,\vec\xi),\HH^G_b(t, \vec\xi')\}
&=& H^G_b(t,\xi) \p_a \delta(\vec\xi
- \vec\xi') - H^G_a(t,\xi') \p_b' \delta(\vec\xi - \vec\xi')\, .
\een
Using this and the fact that $\Pi$ and $T$ have vanishing Poisson bracket
with $H^G_a$ and $H^G_\uparrow$,
we get
\ben \label{egr16}
\{ \HH_\uparrow(t,\vec\xi), \HH_\uparrow(t,\vec\xi')\} &=&  0\, ,
\nonumber \\
\{\HH_\uparrow(t,\vec\xi), \HH_a(t, \vec\xi')\}
&=& \HH_\uparrow(t,\xi') \p_a
\delta(\vec\xi -\vec\xi')\, ,
\een
and
\be \label{egr18}
\{ \HH_a(t,\vec\xi),\HH_b(t, \vec\xi')\} = H_b(t,\xi) \p_a \delta(\vec\xi
- \vec\xi') - H_a(t,\xi') \p_b' \delta(\vec\xi - \vec\xi')\, .
\ee
Eqs.\refb{egr16}, \refb{egr18} generate
the algebra of constraints.

\subsection{Wheeler-de Witt equation in the presence of tachyon field}

We shall now proceed with the
assumption that the `late time' quantum dynamics of the system will be
described by implementing \refb{egr11} and \refb{eexcon} (or \refb{egr14})
in the quantum theory.\footnote{Of course, since the eigenstates of complete
$\HH^T_\perp$ are given by superpositions of positive and negative
$\Pi$ eigenstates, it is not true that the solutions of eqs.\refb{enewcon}
and \refb{egr19} or \refb{egrr19} will also be solutions of corresponding
constraint equations when we use the full $\HH^T_\perp$ given in
\refb{egr9} without the $\Pi>0$ restriction. What we are assuming
here is that eqs.\refb{enewcon} - \refb{egrr19} give a quantum theory which
is approximately {\it equivalent} to the quantum theory associated with full
$\HH^T_\perp$. This is in the same spirit in which we can get a one to one
correspondence between the eigenstates of a quantum mechanical
operator $\wh{\sqrt{\Pi^2 +
V^2}}$
without any restriction on $\Pi$ with that of $\wh\Pi$ under the restriction
$\Pi>0$, both operators having a continuous spectrum of positive
eigenvalues, and having approximately the same density of states at
low energy.} Since these
constraints take identical form to those discussed in
\cite{9409001} for non-rotating, incoherent dust,
we can proceed exactly as in \cite{9409001} to analyze the Wheeler - de
Witt equation.
In particular,
the constraint \refb{egr11},
applied on the wave-functional $\Psi[T(\vec\xi),h_{ab}(\vec\xi)]$,
takes the
form:
\be \label{enewcon}
i \p_a T \, {\delta\Psi\over \delta T(\vec \xi)} =
\wh \HH^G_a(\vec\xi) \, \, \Psi\, ,
\ee
where the hat on top of a variable indicates that it is regarded as an
operator.
On the other hand,
the constraint \refb{egr14}
gives
rise to the `many
fingured time Schrodinger equation' with $T(\vec \xi)$ interpreted as
the time coordinate:
\be \label{egr19}
i {\delta \Psi\over \delta T(\vec\xi)} =  \wh H^G_\uparrow(\vec \xi)
\, \, \Psi\, .
\ee
Note that in order that $\wh H^G_\uparrow(\vec \xi)$
be well defined, we need to restrict the allowed space of
functionals $\Psi[\{T(\vec\xi)\}, \{ h_{ab}(\vec\xi)\}]$ to be in the
subspace spanned by the eigenvectors of $\big(\HH^G_\perp(\vec\xi))^2 -
h^{ab}(\vec\xi) \HH^G_a(\vec\xi)
\HH^G_b(\vec\xi)\big)$ with
positive
eigenvalue. Various subtleties arising from this restriction
have been discussed in
\cite{9409001} and we shall not address these issues here.

We can also use \refb{eexcon} instead of
\refb{egr14} to replace
\refb{egr19} by
\be \label{egrr19}
i {\delta \Psi\over \delta T(\vec\xi)} =  \wh{\left[\Big(1 + h^{ab}(\vec
\xi) \p_a T(\vec \xi) \p_b T(\vec \xi)\Big)^{-1/2} \,
\HH^G_\perp(\vec\xi)\right]}\, \, \Psi\, .
\ee
In this case we require $\Psi[\{T(\vec\xi)\}, \{ h_{ab}(\vec\xi)\}]$ to be
in the subspace of negative $\HH^G_\perp(\vec\xi)$
eigenvalue. Eq.\refb{egrr19} then automatically projects $\Psi$ into
states of positive $\Pi(\vec\xi)$ eigenvalue.
As discussed in
\cite{9409001}, it is not clear whether
\refb{egr19} and
\refb{egrr19} lead to equivalent quantization.

\subsection{Recovering the physics without tachyon field}

In the tachyon field had been absent, the constraints of
general relativity take the
form:
\be \label{em1}
\wh \HH^G_a(\vec\xi) \, \, \Psi
= 0\, , \qquad \wh H^G_\uparrow(\vec\xi)
\, \, \Psi =0\, ,
\ee
or in a  more conventional form:
\be \label{econv1}
\wh \HH^G_a(\vec\xi) \, \, \Psi
= 0\, , \qquad \wh H^G_\perp(\vec\xi)
\, \, \Psi =0\, .
\ee
Given any solution of the constraints \refb{em1}, we can regard this
as a $T$-independent
solution of \refb{egr19},
\refb{enewcon}. On the other hand, any solution of the constraints
\refb{econv1} can be regarded as a $T$-independent solution of
\refb{enewcon}, \refb{egrr19}. This allows us to recover the theory
without tachyon field as a
subsector of the theory with tachyon field. Note that this does not happen
when we
couple a conventional free scalar field to gravity, since even
for a free scalar
field, the energy-momentum tensor
receives contribution not only from the momentum conjugate to
the scalar, but also from the spatial gradient terms and the mass
terms for the
scalar field. Thus acting on an initial wave-functional that is
independent of the scalar field,
the contribution from the momentum term vanishes, but both
the mass$^2$ term as well as the spatial
gradient terms will modify the Wheeler-de Witt equation.
(In perturbation theory
this
is reflected in the fact that even if we begin with a state
containing
only gravitons, their
scattering can pair produce quanta of this scalar field.)

The decoupling of the tachyon at late `time' is consistent with the fact
that the tachyon does not give rise to a physical
particle\cite{0204143,soliton,9909062,phys}, {\it i.e.}
in the scattering of gravitons we should not be able to produce
`quanta' of the tachyon field.\footnote{This
in turn is a consequence of the
fact
that in this approximation the Hamiltonian involving the tachyon field,
including the
terms describing graviton-tachyon interaction, has vanishing
matrix element between $\Pi(\vec\xi)=0$ states.}
Of course this decoupling of the tachyon
occurs only for large $T$ where the
potential term can be ignored. Thus the correct
statement will be that given a
solution of the Wheeler - de Witt equation without the tachyon field, it
can be
regarded
as a $T$-independent solution of the
Wheeler de Witt equation in the presence of
tachyon field for large $T$.
For finite $T$ the solution needs to be modified to
include the effect of the potential term.
Thus all solutions of the Wheeler - de
Witt equation will have a $T$-dependence at early time, showing that
`time' dependent solutions are inevitable in this theory.

\subsection{Gauge fixing}

We can simplify the hamiltonian constraint by `choosing a gauge'
\be \label{eg1}
\p_a T(\vec\xi) = 0\, .
\ee
In concrete terms this means that we choose to look at the wave-functional
$\Psi[\{T(\vec\xi)\},
\{h_{ab}(\vec\xi)\}]$ only over the slice of the
configuration space for which
$T(\vec\xi)=T$ is independent of $\vec\xi$. On such a slice,
\be \label{eg2}
\p_a' \wh T(\vec \xi') \, \Psi[T, \{h_{ab}(\vec{\xi})\}] = 0\, .
\ee
Thus the momentum constraints \refb{enewcon} takes the form:
\be \label{eg3}
\wh \HH^G_a(\vec \xi') \, \Psi[T, \{h_{ab}(\vec{\xi})\}] = 0\, ,
\ee
whereas the hamiltonian constraint \refb{egr19} or \refb{egrr19} take the
form, respectively,\footnote{Formally, once \refb{eg3} is
satisfied, we have $H^G_\perp(\vec \xi)\, \Psi =  H^G_\uparrow(\vec
\xi)\,
\Psi$, and hence \refb{eg4x} and \refb{eg4} give the same equations.}
\be \label{eg4x}
i{\p \over \p T} \, \Psi[T, \{h_{ab}(\vec{\xi})\}]
=\int d^p\xi' \,
\HH^G_\uparrow(\vec\xi')
\, \, \Psi[T, \{h_{ab}(\vec{\xi}\})]\, ,
\ee
or
\be \label{eg4}
i{\p \over \p T} \, \Psi[T, \{h_{ab}(\vec{\xi})\}]
=\int d^p\xi' \,
\HH^G_\perp(\vec\xi')
\, \, \Psi[T, \{h_{ab}(\vec{\xi}\})]\, .
\ee
This has the form of usual Schrodinger equation, with $T$ interpreted as
time. Note however that this form of the equation is modified at small or
finite $T$ where the potential term $V(T)$ is important.

\sectiono{Discussion}  \label{s4}

In this paper we have analyzed the scalar Born-Infeld theory coupled to
gravity, described by the action \refb{egr1}, and shown that at `late
time' the field $T$ appearing in this action could serve as a satisfactory
definition of time in canonical quantum gravity. Since the action
\refb{egr1} in a fixed background metric is able to reproduce some
features
of tachyon dynamics in string theory, our hope is that the field $T$
could arise as some collective mode in open string field theory, and
could lead to
an identification of an intrinsic time variable in string theory.
Clearly much work remains to be done before we have a concrete realization
of this idea.

We shall end this paper by commenting on few related issues.

\begin{enumerate}
\item
In section \ref{s2} we have used a continuity argument to show
that in the limit of large $T$, we can
restrict $\p_0 T$ and hence $\Pi$ to be positive in the whole region of
space-time at late time.
This argument of
course breaks down when we take into account
the effect of $V(T)$ and use the
complete equation \refb{ex5p}, since
here $\p_0 T$ can pass from $-1$ to $1$
smoothly as long as $|\Pi|$ passes from a region  $\Pi \gsim V(T)$
to $\Pi \lsim
- V(T)$. We
note however that for large $T$, when $V(T)$ is exponentially small,
this gives a
very small window in the $\Pi$ space. If we want the effective
field theory to be
valid then the region of space in which this transition from
$\p_0 T \ge 1$ to $\p_0
T\le -1$ takes place must have width
of order
unity or more (measured in string
scale) so that $\p_i \p_0 T$ is of order one or less.
Constraining $|\Pi|$ to be
less
than $V(T)$
in this spatial
region
will give rise to large quantum fluctuations in $T$ of order
$e^{\alpha T/2}$ (for
$V(T)=\exp(-\alpha T/2)$) and will invalidate our classical
analysis.

In string theory one can construct configurations where different
regions of space-time have $\Pi$ with different sign as follows.
Consider a pair of non-BPS D0-branes (or two D0-$\bd0$ pair in the
case of type IIA string theory) separated by a distance. On each we
set up a rolling tachyon solution given in \cite{0203265}, but we
adjust one of the solutions to have a large time lag relative to the
other. As a result during a large interval of time, as on one of the
D0-branes the tachyon rolls up the hill, giving rise to negative
$\Pi$, on the other D0-brane the tachyon will roll down the hill and
will have positive $\Pi$. Thus one could ask if quantum fluctuations
around this background in string theory shows any unusual divergence
of the kind predicted from the field theory analysis. Since the one
loop partition function in the presence of this D-brane system
involves computing the inner product of the boundary states
associated with these two D0-branes, we see that if the boundary
states had been finite in the $x^0\to\pm\infty$ limit,
we would not expect any unusual behaviour of
the partition function. However recent analysis shows that
coefficients of some terms in the boundary state grows exponentially
with time\cite{0208196,0208142}. It is tempting to speculate that
this exponential growth is related to the exponentially large quantum
fluctuations that we expect from such a configuration in the
effective field theory.

\item
One question one could ask is whether it is possible to say something
concrete about the quantum theory described by the action \refb{ei1} even
before  we couple this to gravity. Of course this is not a standard
renormalizable field theory, but we could in principle use some kind of
lattice regularization, and treat this as a theory with a finite cut-off.
To begin with we can consider a simpler model, --
that for $p=0$, -- for which the
Hamiltonian takes the form:
\be \label{es1}
\wh H = \wh{\sqrt{\Pi^2 + (V(T))^2}}\, .
\ee
Since $\wh\Pi^2 + \wh V^2$ is a positive definite hermitian operator,
we can define $\wh H$ as the positive square root of $\wh\Pi^2 + \wh
V^2$. $(\wh\Pi^2 + \wh V^2)$ for $\wh V=e^{-\alpha \wh T/2}$ has a
continuous spectrum beginning at 0. An eigenstate of $(\wh\Pi^2 + \wh
V^2)$ with eigenvalue $\lambda^2$ has the asymptotic form
proportional to $\sin(\lambda T + \phi(\lambda))$ where
$\phi(\lambda)$ is a phase factor.  There is, however, no strictly
zero energy state, since $\phi(\lambda)\to 0$ as $\lambda\to 0$.

In order to analyze the time evolution of a state under the
Hamiltonian \refb{es1}, we can decompose it as a linear combination
of the eigenstates of $(\wh\Pi^2 + \wh V^2)$, and multiply the
coefficient of the eigenstate with eigenvalue $\lambda^2$ by
$e^{i\lambda t}$. We shall not do this explicitly here, but from
general considerations we expect that if we begin with a gaussian
wave-packet, and evolve it according to this scheme, then in the far
future the wave-function, {\it when decomposed in the basis of plane
waves on an infinite line}, will consist mostly of right-moving
waves, since all the left-moving components will get reflected by the
tachyon potential.

If we
begin with some wave-function that approaches a constant value at
$\infty$,
then the Schrodinger
equation
\be \label{es2}
i {\p \Psi(T;x^0)\over \p x^0} = \sqrt{\wh\Pi^2 + \wh V^2}
\, \Psi(T; x^0)\, ,
\ee
has the property that for large $T$,
the right hand side of the equation is
small and hence
the wave-function remains constant. However the wave-function does
evolve
with time in a non-trivial manner for $T\lsim 1$,
since the true ground state for which the
right hand side of \refb{es2} vanishes is localized at $T=\infty$.

For the case of general $p$ (including the space-filling branes)
we can define the
Hamiltonian density in a similar way, by taking this to be
the positive square root
of the operator:
\be \label{es3}
(1 + \wh {\dt}) (\wh\Pi^2 + \wh V^2)\, .
\ee
Note that $[\wh \Pi(x), \p_i \wh T(x)]
\propto \delta'(0)=0$. To see this
in
a
properly
regularized theory we can use a lattice regularization with
lattice sites labelled
by $s$, and
identify $\wh\Pi(x)$ with $\wh\Pi_s(x^0)$ and $\p_i \wh T(x)$ with
$(\wh T_{s+1}(x^0) -
\wh T_{s-1}(x^0))$ where we are using a shorthand notation
where $\wh T_{s\pm 1}$
denotes the
lattice site displaced
$\pm 1$ unit from the site $s$ along the $i$th direction.
$\wh\Pi_s(t)$ clearly commutes with $(\wh T_{s+1}(x^0) -
\wh T_{s-1}(x^0))$. Thus there is no operator ordering problem in
defining
\refb{es3} at individual sites
and it is a positive definite hermitian operator.
This allows us to define its positive definite square root.
However this analysis also shows that the hamiltonian densities at
different sites do not commute, and hence are not simultaneously
diagonalizable. This makes the concrete analysis of the spectrum difficult.
Nevertheless this definition of the
Hamiltonian allows us to calculate the action of
the hamiltonian on a give wave-function
explicitly. For example, in the lattice
version, in order to calculate
$\wh\HH_s\Psi$ for a given wave-functional
$\Psi(T_1,\ldots T_n)$, we first multiply $\Psi$ by the lattice version of
$\sqrt{1+\dt}$ associated with the
$s$-th site, then express the result as a linear
combination of the eigenfunctions
of $(\wh \Pi_s^2 + (V(\wh T_s))^2)$, and finally
multiply each term in the expansion
by the positive square root of the associated
eigenvalue. We can repeat this for
each site $s$, and the final action of the full
Hamiltonian on $\Psi$ is obtained by
adding up the contribution from each lattice
site.
We can then find the time evolution of the state by solving
the time dependent Schrodinger
equation:
\ben \label{es4}
i \, {\p \Psi(T(\vec x);x^0)\over \p x^0} &=& \int d^p x' \,
\wh{\sqrt{(1 +
{\dt}) \, (\Pi^2 +
V^2)}} \, \,
\Psi(T(\vec x); x^0)
\nonumber \\
&=& \sum_s \sqrt{(1 + (\vec\nabla T_s)^2) (\Pi_s^2 + V_s^2)}\,
\Psi(T_1,\ldots T_n;x^0)\, ,
\een
where in the last line of the above equation we have the
lattice version of the Hamiltonian, with $\vec\nabla T_s$ defined by
taking appropriate difference between the values of
$T$ at neighbouring
sites. This equation
will have the property that
if we take
an initial wave-function which approaches
a constant for large $\{T(\vec x)\}$,
then it remains constant for large $\{T(\vec
x)\}$ as the wave-function evolves in time.
More precisely a configuration where $\Psi$ is a constant independent of
$\{T(\vec x)\}$ and $x^0$ will satisfy \refb{es4} approximately in the
region $T(\vec x) \ge M$, $|\vec\nabla T|<<e^{\alpha M/2}$ for a
sufficiently large number $M$, but fails in the region where $T(\vec x)$
is finite, or $|\vec\nabla T|$ is of order $e^{\alpha T/2}$.
If we are only interested in studying the properties of the wave-function
for large $T$ and not too large $|\vec\nabla T|$,
then we could treat this as
the `approximate vacuum state',
although strictly it is not an eigenstate of the Hamiltonian.

\end{enumerate}

\medskip

{\bf Acknowledgement}:
I would like to thank
S.~Minwalla, E.~Rabinovici, A.~Strominger and B.~Zwiebach
for useful
discussions.
This work was supported in part by a grant
from the Eberly College
of Science of the Penn State University,
and a grant from the NM Rothschild and
Sons Ltd at the Isaac Newton institute.
I would also like to acknowledge
the hospitality of YITP, Stony Brook,
and the Center for Theoretical Physics at MIT
where part of this work was done.

\end{document}